\documentclass[pdflatex,sn-mathphys-num]{sn-jnl}% Math and Physical Sciences Numbered Reference Style
\usepackage{graphicx}%
\usepackage{multirow}%
\usepackage{amsmath,amssymb,amsfonts}%
\usepackage{amsthm}%
\usepackage{mathrsfs}%
\usepackage[title]{appendix}%
\usepackage{xcolor}%
\usepackage{textcomp}%
\usepackage{manyfoot}%
\usepackage{booktabs}%
\usepackage{algorithm}%
\usepackage{algorithmicx}%
\usepackage{algpseudocode}%
\usepackage{listings}%
\usepackage{colortbl}%
\theoremstyle{thmstyleone}%
\theoremstyle{thmstyletwo}%

\theoremstyle{thmstylethree}%

\newcommand*\tensor[1]{\underline{\mathbf{#1}}}

\begin{document}

\title[THz nonlocal response in graphene]{Terahertz s-SNOM reveals nonlocal nanoscale conductivity of graphene}

%%=============================================================%%
%% GivenName	-> \fnm{Joergen W.}
%% Particle	-> \spfx{van der} -> surname prefix
%% FamilyName	-> \sur{Ploeg}
%% Suffix	-> \sfx{IV}
%% \author*[1,2]{\fnm{Joergen W.} \spfx{van der} \sur{Ploeg} 
%%  \sfx{IV}}\email{iauthor@gmail.com}
%%=============================================================%%

\author[1,3]{\fnm{Henrik Bødker} \sur{Lassen}}

\author[1]{\fnm{William Vang} \sur{Carstensen}}

\author[2]{\fnm{Leonid} \sur{Iliyshyn}}

\author[2]{\fnm{Timothy J.} \sur{Booth}}

\author[2]{\fnm{Peter} \sur{Bøggild}}

\author*[1]{\fnm{Edmund John Railton} \sur{Kelleher}}\email{edkel@dtu.dk}

\author*[1]{\fnm{Peter Uhd} \sur{Jepsen}}\email{puje@dtu.dk}

\affil*[1]{\orgdiv{Department of Electrical and Photonics Engineering}, \orgname{Technical University of Denmark}, \city{Kongens Lyngby}, \postcode{DK-2800}, \country{Denmark}}

\affil[2]{\orgdiv{Department of Physics}, \orgname{Technical University of Denmark}, \city{Kongens Lyngby}, \postcode{DK-2800}, \country{Denmark}}

\affil[3]{\orgdiv{Current address: Department of Physics, Faculty of Science}, \orgname{Kyoto University}, \orgaddress{\city{Kyoto}, \postcode{606-8502}, \country{Japan}}}

\abstract{As photonic and electronic technologies approach nanometre length scales and terahertz operating speeds, electrical conductivity can no longer be treated as a purely local material parameter. In this regime, charge transport becomes intrinsically nonlocal, with conductivity depending on both frequency and momentum, $\sigma(\omega,q)$, fundamentally limiting field confinement, dispersion, and loss in nanoscale devices. Here, we directly measure the nonlocal nanoscale conductivity of graphene using terahertz scattering-type near-field optical microscopy. By combining broadband THz near-field spectroscopy with quantitative electrodynamic modelling, we extract the complex conductivity of single- and few-layer graphene with $\sim$50~nm spatial resolution. 
We find that nonlocal response dominates the terahertz conductivity of monolayer graphene even at length scales comparable to practical device dimensions. 
These results establish nonlocal conductivity as a measurable and design-relevant material property in the terahertz regime, providing a quantitative foundation for predicting performance limits in ultracompact photonic and electronic systems. 
}


%%\pacs[JEL Classification]{D8, H51}

%%\pacs[MSC Classification]{35A01, 65L10, 65L12, 65L20, 65L70}

\maketitle

\section{Introduction}

As photonic and electronic technologies approach nanometre dimensions and terahertz (THz) operating speeds, the electrical conductivity of materials can no longer be treated as a purely local quantity. In low-dimensional conductors, such as nanowires, nanotubes, and atomically thin crystals, charge transport is governed by quantum and collective effects that emerge on length scales comparable to device dimensions, rendering classical Ohmic transport insufficient~\cite{Voit1995, Bockrath1999, Cai2025}. In this regime, conductivity becomes intrinsically dispersive in both frequency and momentum, $\sigma(\omega,q)$, and spatially varying electric fields probe electronic dynamics beyond the local-response approximation. Direct access to conductivity with nanometre spatial resolution is therefore essential for both understanding light-matter interaction \cite{Huber2008, Lundeberg2017, Woessner2015, Fei2011}, enabling a quantitative validation of nonlocal transport models~\cite{Lovat2013, Falkovsky2007}, and establishing predictive and physically grounded design rules, confinement limits and loss mechanisms for nanoscale photonic and electronic systems, where nonlocal and quantum transport effects fundamentally modify device scaling and performance \cite{Mortensen2014, Basov2016}.

Nonlocal electrical response has long been anticipated by foundational theories of the electron gas and linear response~\cite{Lindhard1954, Mermin1970, Kubo1957}, yet direct experimental access to the nonlocal conductivity, $\sigma(\omega,q)$, in real materials has remained limited. Conventional contact-based transport measurements, even on nanowires and quantum dots, yield device-integrated conductance that reflects terminal-to-terminal transmission rather than spatially resolved carrier dynamics~\cite{Buttiker1986, Beenakker1991, Badawy2024}. Optical probes typically access regimes where nonlocal effects are strongly suppressed. As a result, nonlocality is often treated as a small correction in nanophotonic modelling, despite the fact that deeply subwavelength confinement intrinsically involves electronic length scales. Optical probes typically access regimes where nonlocal effects are strongly suppressed~\cite{Mortensen2014, Raza2015, Basov2016}. As a result, nonlocality is often treated as a small correction in nanophotonic modelling, despite the fact that deeply subwavelength confinement intrinsically involves electronic length scales. 

Scanning near-field optical microscopy (SNOM) overcomes the diffraction limit by exploiting evanescent fields localised at a sharp metallic tip, enabling optical access to large in-plane momenta~\cite{Ash1972, Keilmann2004}. Quantitative scattering models allow the complex local response to be retrieved spectroscopically~\cite{Cvitkovic2007}, establishing near-field optics as a powerful probe of nanoscale electrodynamics. Recent near-field experiments have resolved THz conductivity at the nanoscale~\cite{Jing2021}, visualised ultrafast phase transitions in two-dimensional materials~\cite{Siday2022}, and tracked propagating graphene plasmons in space and time~\cite{Anglhuber2025}.

However, these same experiments expose a fundamental limitation of the local-response picture. When electromagnetic fields are confined to nanometre scales, the assumption of a purely local conductivity fails. 
In this regime, $\sigma(\omega,q)$ acquires a pronounced momentum dependence, leading to intrinsic limits on field confinement, dispersion, and quality factor. Such nonlocal effects have recently been inferred in ultrasmall THz plasmonic cavities~\cite{Aupiais2023} and slow-light graphene plasmon systems~\cite{Lundeberg2017}.

The terahertz frequency range provides a uniquely favourable window for directly probing nonlocal electronic response. At THz frequencies, long carrier mean free paths and deeply subwavelength near-field confinement combine to produce electric fields that vary on length scales comparable to electronic diffusion lengths. 
In contrast to mid-infrared or optical frequencies, where nonlocality is confined to a few nanometres, the THz regime amplifies nonlocal effects to tens or hundreds of nanometres, rendering them experimentally accessible.

Here, we directly measure the nonlocal nanoscale conductivity of graphene using terahertz scattering-type near-field optical microscopy. By combining broadband THz near-field spectroscopy with quantitative electrodynamic modelling, we extract the complex conductivity, $\sigma(\omega,q)$, of single- and few-layer graphene with $\sim$50~nm spatial resolution. We find that nonlocal response dominates the THz conductivity of monolayer graphene even at length scales comparable to practical device dimensions. Hence, our results establish nonlocal conductivity as a measurable, design-relevant material property in the THz regime rather than a small theoretical correction. 

%% in Nature Photonics, the Results section does not have to have the name "Results", it can be broken down into several sections with descriptive titles

\section{Near-field THz mapping of exfoliated graphene}\label{sec:mapping}

\begin{figure}[ht]
\centering
\includegraphics[width=0.7\textwidth]{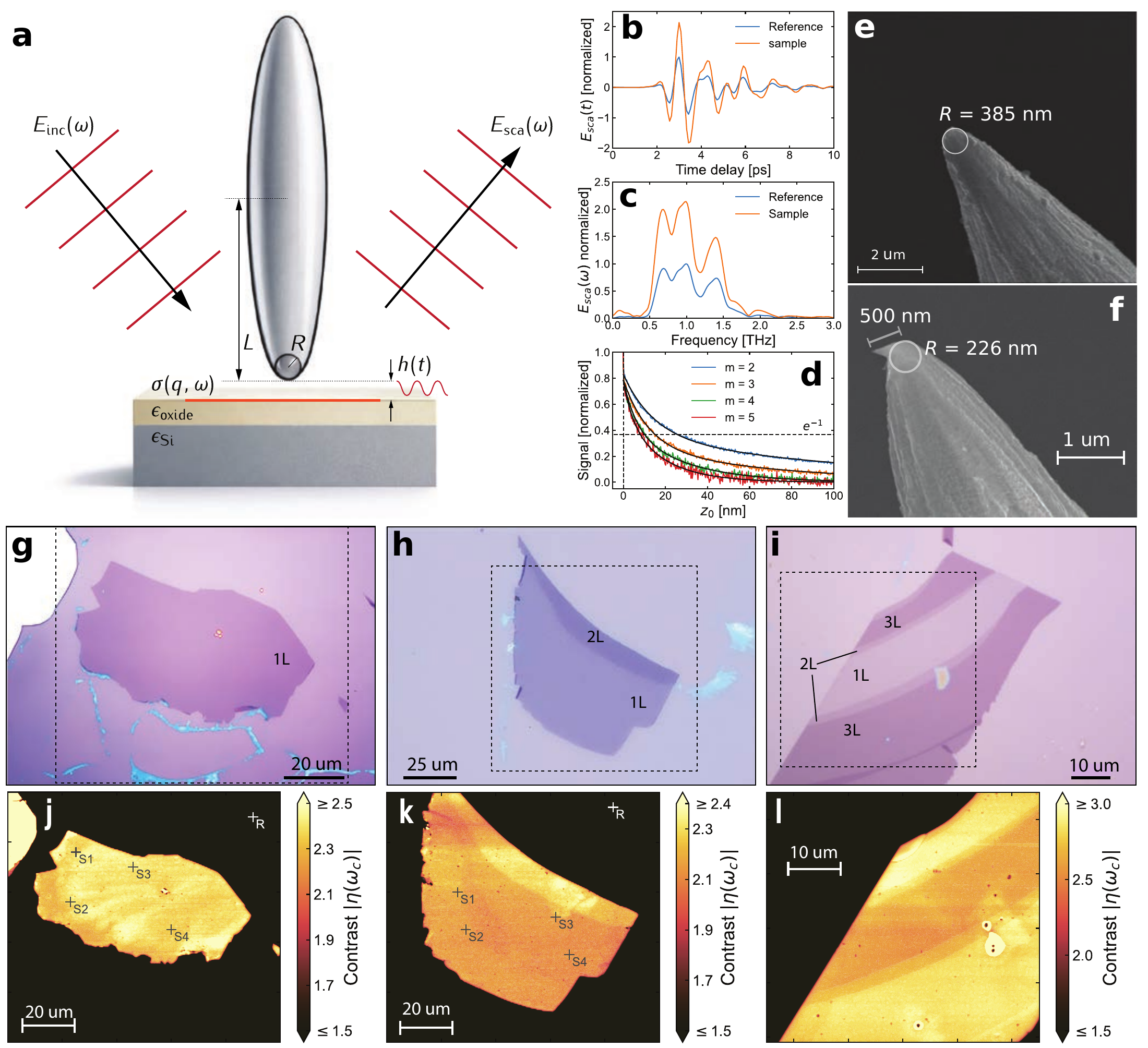}
\caption{$|$ \textbf{THz-frequency scattering-type Scanning Near-Field Optical Microscopy (THz-SNOM) instrumentation and imaging of exfoliated graphene flakes.}
\textbf{a.} Schematic of the tip-sample interaction, featuring a tapping-mode AFM adapted for in- and out-coupling of single-cycle THz pulses to a metallic tip.
\textbf{b.} Time-domain waveforms of the scattered THz pulse at the tip apex (orange) versus substrate reference (blue) (\textbf{b}). \textbf{c.} Corresponding spectral amplitudes spanning 0.5-1.5~THz.
\textbf{d.} Tip–sample approach curves for demodulation orders m = 2–5, showing enhanced near-field confinement at higher harmonics; m = 2 (grey) is used for all subsequent measurements for optimal signal-to-noise ratio.
\textbf{e,f.} Representative SEM images of dented AFM tips, revealing radii of $\sim$385~nm and non-spherical apex shapes with $\sim$226~nm radius, produced by gentle denting to boost the scattering signal.
\textbf{g,h,i.} Optical micrographs of mechanically exfoliated graphene on SiO$_2$/Si for samples S1-S3: monolayer (1L) in S1; predominantly monolayer with a narrow bilayer (2L) region in S2; and a central monolayer region flanked by bi- and trilayer (3L) regions in S3. Polymer residues (bright blue) and bulk graphite (bright region, top left of \textbf{g}) are also visible.
\textbf{j,k,l.} White-light mode THz-SNOM contrast images at $\omega_c/2\pi=1.2\text{~THz}$ for S1, S2 and S3, recorded with the temporal delay parked at the substrate peak; distinct layer-dependent contrast and intralayer variations are clearly resolved.}
\label{fig:intro_figure}
\end{figure}

THz-frequency scattering-type Scanning Near-Field Optical Microscopy (THz-SNOM) measurements are performed with a commercial instrument (Attocube THz-NeaSCOPE) as shown schematically in Fig.~\ref{fig:intro_figure}(a). The core of the instrument is a tapping-mode Atomic Force Microscope (AFM) adapted to accommodate in- and out-coupling of single-cycle THz-frequency electromagnetic pulses (THz pulses) interacting with the metallic AFM tip. Overtones of the scattered signal are detected by demodulation techniques, giving access to the amplitude and phase of the near-field component of the scattered field~\cite{Hillenbrand2000}. Temporal and spectral waveforms of scattered THz pulses are shown in Fig.~\ref{fig:intro_figure}(b) and (c), showing a useful spectral coverage spanning 0.5--1.5~THz. Approach curves, recorded in tapping mode by varying the minimal sample-tip distance, are shown in Fig.~\ref{fig:intro_figure}(d) for demodulation orders $m=2-5$, indicating the expected increasing confinement at higher demodulation orders. For the best signal-to-noise ratio (SNR), we will use the $m=2$ demodulation order in the following. 

The nominal radii of the pristine AFM tips are below 50~nm, and after gentle denting, we increase the tip radii to several hundred nanometres, as shown in the representative scanning electron microscopy (SEM) images of dented AFM tips in Fig.~\ref{fig:intro_figure}(e) and (f). The denting serves to increase the signal strength~\cite{Kramer1996, Knoll1997}. Denting can lead to non-spherical apex shapes, as shown in Fig.~\ref{fig:intro_figure}(f), which must be considered in the quantitative model-based interpretation of results.

We utilise the THz-SNOM system to investigate mechanically exfoliated graphene flakes with lateral dimensions of order 100~$\mu$m. High-contrast microscope images~\cite{Jessen2018} of the selected samples are shown in Fig.~\ref{fig:intro_figure}(g,h,i). The first sample (S1) is monolayer (1L) graphene, the second sample (S2) features a large single-layer region and a narrow bilayer (2L) region. The third sample (S3) features a central monolayer region flanked by bilayer and trilayer (3L) regions. The microscope images also reveal polymer residues (bright blue) and graphite (bright region, top left corner of Fig.~\ref{fig:intro_figure}(g)).

THz-SNOM images of the contrast between the SiO$_2$/Si substrate and the graphene flakes are shown in Fig.~\ref{fig:intro_figure}(j,k,l). The images are recorded in the so-called white-light (WL) mode, where the temporal delay is parked at the peak of the THz signal recorded on the substrate (orange trace in Fig.~\ref{fig:intro_figure}(b)). At this temporal position, all frequency components of the THz signal contribute constructively, and the image can be considered a spectral average of the response. The weighted frequency is $\omega_c/2\pi=1.2\text{~THz}$. We observe clear variation in the scattering signal contrast, $\eta(\omega_c)$, for the monolayer sample, with a pattern visually resembling eroded geological features in a landscape map. Sample S2 displays a monolayer region with fewer features than S1, and a distinctly identifiable bilayer region with a higher scattering signal and some intralayer contrast variations. The THz-SNOM image of sample S3 clearly identifies the mono-, bi and trilayer regions, and we again observe variations within the individual layer regions. 
The images demonstrate that, with sufficient SNR, THz-SNOM can identify layer count and response variations within each layer.     

\section{THz spectroscopic evidence of nonlocal conductivity response}\label{sec:spectroscopy}

\begin{figure}[h]
\centering
\includegraphics[width=0.7\textwidth]{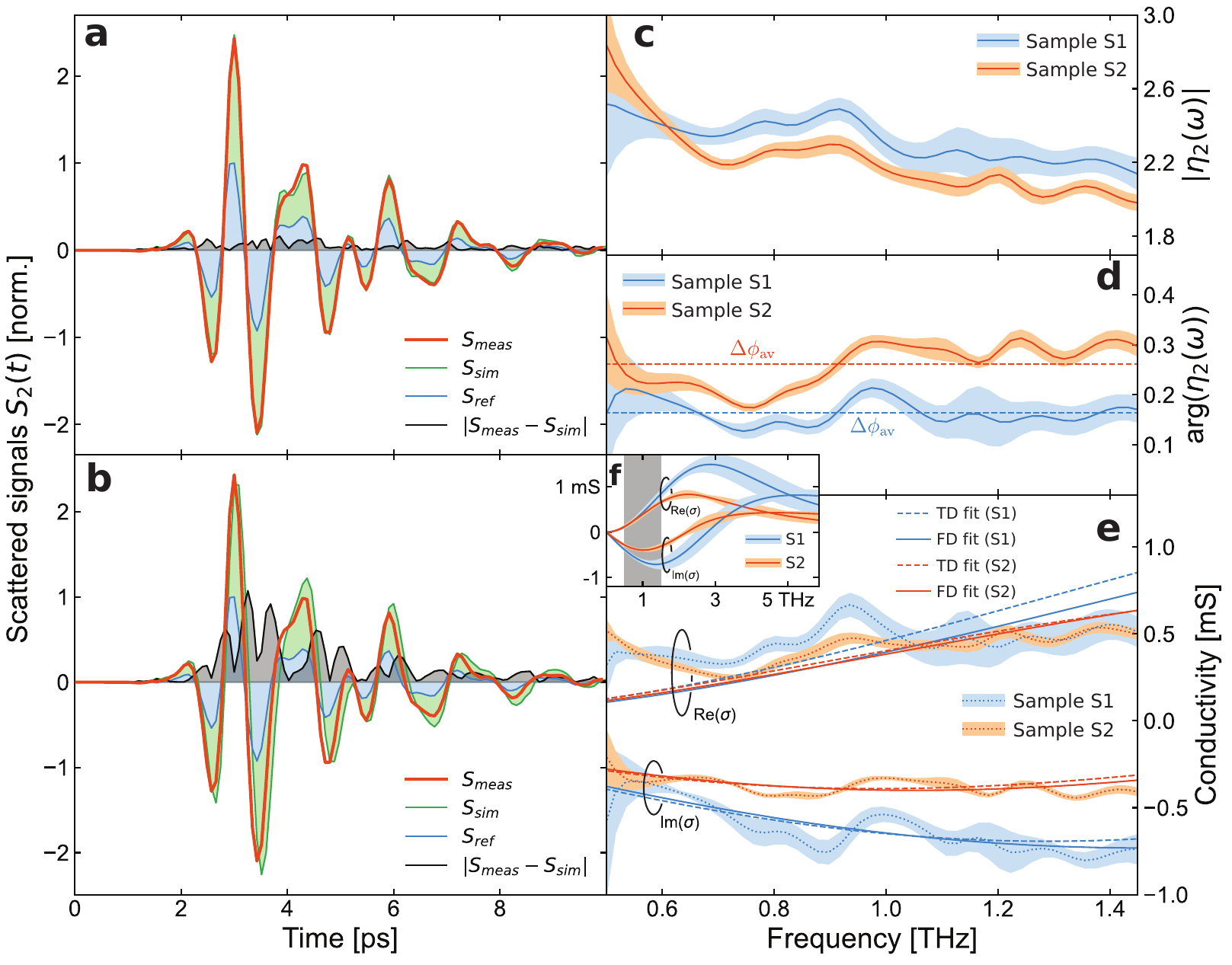}
\caption{$|$ \textbf{THz-SNOM spectroscopy of monolayer graphene conductivity.}
\textbf{a.} Time-domain THz waveforms recorded on the bare substrate at reference position R (blue) and on graphene sample S1 at position P2 (red), overlaid with the finite-dipole model (FDM) simulation using the nonlocal Lovat conductivity (green) and the absolute difference $|S_{\rm meas}-S_{\rm sim}|$ (grey).
\textbf{b.} Same experimental data fitted with a local Drude-like conductivity, showing markedly poorer agreement.
\textbf{c,d.} Spectrally resolved amplitude (\textbf{c}) and phase (\textbf{d}) contrasts, respectively, averaged over four graphene positions (P1-P4) relative to references for samples S1 (blue) and S2 (red); shaded regions denote $\pm$1 standard deviation between measurements.
\textbf{e.} Real and imaginary parts of the complex conductivity recovered via inversion of the FDM for both samples. Solid curves show Lovat-model fits to the frequency-domain spectra, while dashed curves are the model conductivities obtained from the time-domain minimisation in \textbf{a}.
\textbf{f.} Inset: Lovat-model conductivity spectra extended to 0.1-5.5~THz; grey shading indicates the spectral coverage of the instrument, and envelopes represent the fit standard deviation between points P1-P4.}
\label{fig:conductivity_spectra}
\end{figure}

The position markings in Fig.~\ref{fig:intro_figure}(j,k) identify locations where single-point, full-waveform scans were performed, allowing spectrally resolved measurements. We use these point measurements to identify the nonlocal nature of the graphene conductivity and compare the measured conductivity spectra with a nonlocal conductivity model of graphene based on semiclassical Boltzmann transport theory under the relaxation approximation and the Bhatnagar-Gross-Krook (BGK) model, analytically formulated by Lovat et al.~\cite{Lovat2013} and the conventional Drude-like conductivity~\cite{Falkovsky2007} emerging from the local limit of the Kubo formalism~\cite{Kubo1957}. Figure~\ref{fig:conductivity_spectra} shows a summary of our observations based on spectroscopy of the frequency-dependent conductivity. 

Figure~\ref{fig:conductivity_spectra}(a) shows THz temporal waveforms recorded on the substrate at position R (blue curve, $S_\text{ref}$) and at position P2 (red curve, $s_\text{meas}$) on sample S1. The green curve ($S_\text{sim}$) is a simulated response, based on the Finite Dipole Model (FDM) with a layered sample structure and using the Lovat-BGK formulation of the nonlocal conductivity of graphene, with the absolute difference between the measured and simulated waveforms shown as the grey curve ($|S_\text{meas}-S_\text{sim}|$). Figure~\ref{fig:conductivity_spectra}(b) shows the same experimental data compared to the best fit obtained using local Drude-like conductivity, with a significantly lower quality of the fit. 

The spectrally resolved amplitude and phase of the contrast between the scattered signal averaged over the four sample positions P1-P4 and the reference positions R of the two samples S1 and S2, together with the standard deviation between measurements on each sample, is shown in Fig.~\ref{fig:conductivity_spectra}(c) and (d), respectively. These spectra are used as input to an FDM-based numerical analysis to retrieve the complex conductivity spectrum of monolayer graphene, as shown in Fig.~\ref{fig:conductivity_spectra}(e). The real part of the conductivity increases with frequency, while the imaginary part is negative and shows a soft minimum within the measurement spectral range. The solid curves are best fits with the Lovat-BGK analytical model to the spectrally resolved conductivity measurements, while the dashed curves are the model conductivities resulting from the time-domain minimisation shown in Fig.~\ref{fig:conductivity_spectra}(a). Results for samples S1 and S2 are shown in blue and red colours, respectively. The inset (Fig.~\ref{fig:conductivity_spectra}(f)) displays the fitted Lovat-BGK conductivity model spectra and their standard deviation between the fits to the individual points over a wider frequency range (0.1-5.5~THz), offering greater insight into the full spectra of the nonlocal response of the two samples. 
The grey area indicates the spectral coverage of our instrument. 

The nonlocal character of the conductivity response is uniquely identified here; in contrast to our observations, a local Drude-like response with frequency dependence $\propto(1-i\omega\tau)^{-1}$ remains positive at all frequencies, with decreasing real part and increasing imaginary part for $\omega\tau<1$. The nonlocal response, on the other hand, exhibits behaviour reminiscent of a resonant response, arising from the interplay between carrier inertia and spatially dispersive charge-density dynamics.

\section{Superfocusing and nonlocal response}\label{sec:nonlocality}

The best-fit parameters are detailed in the Extended Data Section. As described in the Methods Section, we performed the time-domain minimisation as the first step in the fitting procedure. A good agreement between the experimental data and the Lovat model in the FDM framework was achieved with a Fermi energy and scattering time of $(231\pm 41)$~meV and $(62\pm 10)$~fs for sample S1 and $(157\pm 9)$~meV and $(50.2\pm 0.5)$~fs for sample S2, averaged over the four point measurements P1-P4. For S1 and S2, the optimal tip radius $R$ used in the FDM was $(313\pm 63)$~nm and $(376\pm 18)$~nm, respectively, and the optimal value of the in-plane momentum $q$ used in the Lovat conductivity expression was $(0.020\pm 0.002)$~nm$^{-1}$ and $(0.019\pm 0.001)$~nm$^{-1}$ for the two samples. These in-plane momenta correspond to a length scale ($d\sim1/q$) of $(50\pm 4)$~nm and $(54\pm 2)$~nm, respectively, at first glance in disagreement with the values obtained for the optimal tip radius. However, as we will discuss now, this apparent disagreement is only superficial and rooted in the superfocusing of the incident THz field under the AFM tip. 

\begin{figure}[h]
\centering
\includegraphics[width=0.7\textwidth]{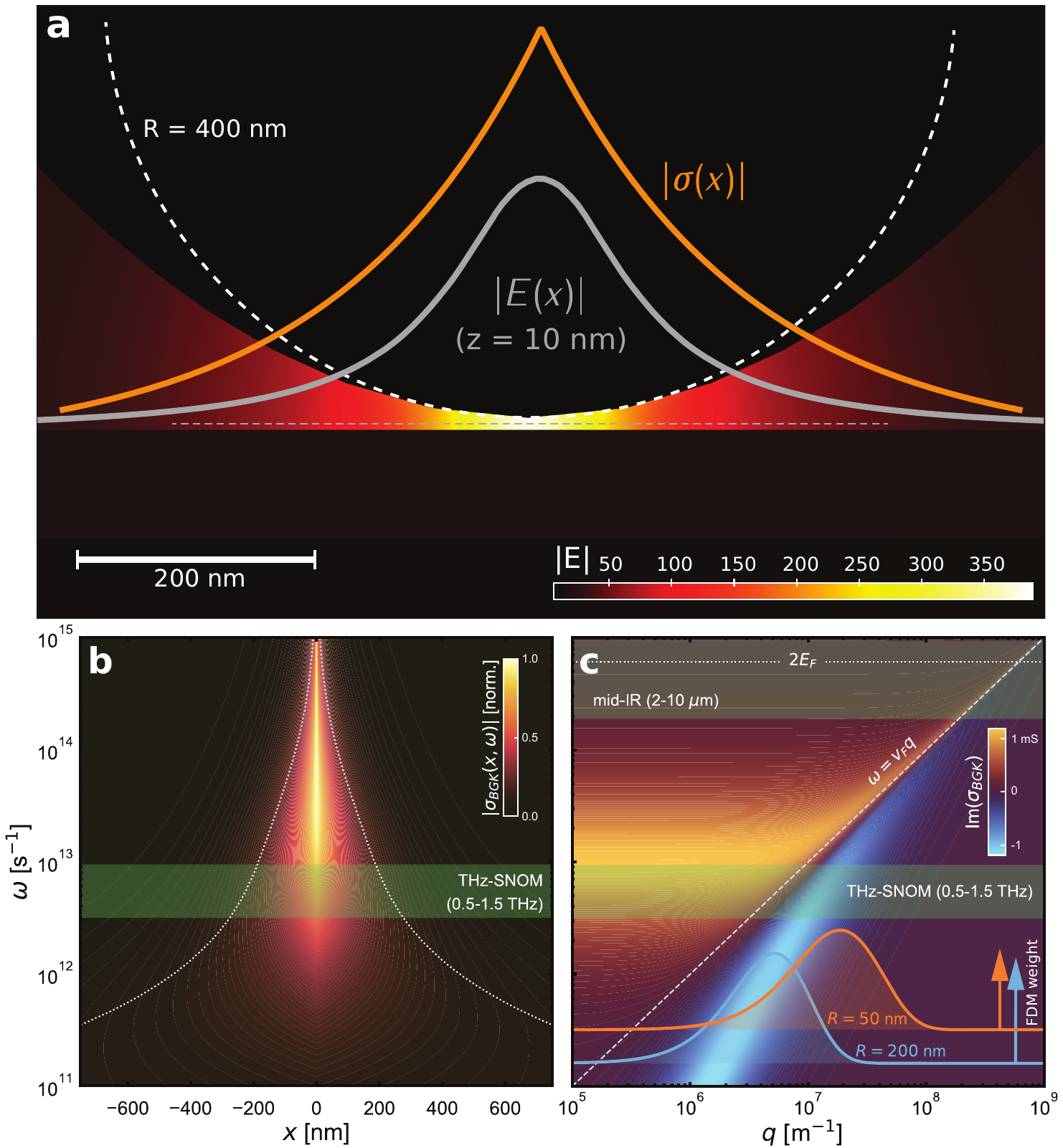}
\caption{$|$ \textbf{Field distribution and nonlocal response in THz‐SNOM.}
\textbf{a.} Finite-element calculation of the near-field enhancement at 1~THz in the apex region of an AFM tip (ellipsoidal cross-section in black) with radius $R = 400\text{~nm}$, positioned 20~nm above the sample surface. The white dashed curve shows the circular approximation near the apex. False-colour shading plots the field enhancement $|E(z = 10~nm)|/E_0$, revealing a confinement with full-width at half-maximum (FWHM) $\approx 90\text{~nm}$ (grey line), much smaller than the tip radius. The real-space nonlocal response of the Lovat-BGK model at $\omega/2\pi = 1\text{~THz}$ ($\tau = 100\text{ fs}, E_\text{F} = 200\text{ meV}$), $|\sigma_\text{BGK}(x)|$ is overlaid in orange, showing that the nonlocal "reach" is comparable to, and even exceeds the extent of the THz field.
\textbf{b.} Spatial profiles of $|\sigma_\text{BGK}(x,\omega)|$ obtained by inverse Fourier transform of $\sigma_\text{BGK}(q,\omega)$ over the frequency range $\omega = 10^{11}–10^{15} s^{-1}$. The dashed white curves trace the FWHM of the nonlocal response at each $\omega$, illustrating a reach of a few hundred nanometres in the THz (green band: 0.5–1.5~THz) that collapses to only a few nanometres in the mid-IR.
\textbf{c.} Imaginary part $\text{Im}(\sigma_\text{BGK}(q,\omega)$ plotted as a function of momentum $q$ and frequency $\omega$. The white dashed line marks $\omega = v_\text{F}q$, below which $\text{Im}(\sigma_\text{BGK})$ becomes negative, a clear signature of nonlocality, while the dotted line indicates the onset of interband transitions at $\hbar\omega = 2E_\text{F}$. The THz-SNOM band is shown in green, and the weighted in-plane momentum distributions (FDM weight) for tip radii $R = 200\text{~nm}$ (blue) and $R = 50\text{~nm}$ (orange) illustrate the $q$-values probed in the experiment and used in the Lovat model evaluation, respectively.}\label{fig:nonlocality}
\end{figure}

Figure~\ref{fig:nonlocality}(a) shows a quantitative model of the apex region of an AFM tip with radius $R=400$~nm, hovering 20~nm over the sample surface. The ellipsoidal cross-section of the tip is shown in black, and the white, dashed curve shows the circular approximation to the shape near the tip apex. 
The colour map reproduces the field enhancement relative to the incident field strength at a frequency of 1~THz, calculated with the Finite Element Method (FEM, see Methods Section for details). The grey curve represents the field strength in the gap halfway between the sample and the tip, evaluated at $z=10$~nm, showing that the field is concentrated within a radius of 90~nm, a radius significantly smaller than the tip radius of 400~nm. A similar simulation of the field localisation under a 50-nm tip radius (not shown here) shows field confinement to a radius of 35~nm. Hence, for large tips, the field confinement is significantly higher than the radius of the tip. For comparison, a numerical evaluation of the nonlocality of the Lovat model with a scattering time $\tau=100\text{ fs}$ and Fermi energy $E_\text{F}=200\text{ meV}$ at 1~THz (inverse Fourier transformed from momentum to real space) is shown with the orange curve in Fig.~\ref{fig:nonlocality}(a), showing that the reach of the nonlocality is comparable to (and even larger than) the extent of the localised THz field.  
As detailed further in Fig.~\ref{fig:nonlocality}(b), the width of the nonlocal response rapidly reduces at shorter wavelengths. 
We have calculated the real-space response of the nonlocal Lovat model over the frequency range $\omega=10^{11}-10^{15}\text{~s}^{-1}$ ($\sigma_\text{BGK}(x,\omega)={\cal{F}}^{-1}(\sigma_\text{BGK}(q,\omega)$). The dashed curves indicate the FWHM of the spatial extent of the nonlocal response. At high frequencies, corresponding to the mid-IR, we see a reach of the nonlocal response of only a few nm, while in the THz range, as exemplified also in Fig.~\ref{fig:nonlocality}(a), the nonlocality has an extent of a few hundred nm, larger than the extent of the localised THz field in the THz-SNOM apparatus. Hence, the low probe frequency with a high in-plane momentum in the deep sub-wavelength limit, only accessible with SNOM instrumentation at THz frequencies or lower, is the right combination for clear and direct observation of the nonlocal response of the conductive electrons in graphene and other conductive systems. 
Figure~\ref{fig:nonlocality}(c) shows an overview of the imaginary part of the nonlocal conductivity response of graphene. The imaginary part shows most clearly the signature of nonlocality, in the sense that for frequencies $\omega<v_\text{F}q$, $\text{Im}(\sigma_\text{BGK})$ displays distinct, negative values, in contrast to conventional, local response where $\text{Im}(\sigma)>0$. With the THz-SNOM system, we access exactly this signature region of the nonlocal response, indicated by the THz-SNOM frequency range (green, horizontal bar) and the weighted distribution of probed in-plane momenta shown as orange and blue curves, calculated for tip radii $R=50, 200\text{~nm}$, respectively. The in-plane momentum weight is detailed in the Methods section. In contrast, a similar investigation carried out at higher frequencies, for instance in the mid-IR approaching the transition at $2E_\text{F}$ for the onset of interband transitions, would access the local response of the system.  
  
\section{Direct mapping of conductivity, Fermi energy and scattering time}\label{sec:fermi_tau}

Based on the frequency-resolved spectra of the nonlocal conductivity measured at specific points (P1-P4) on the samples, we can use the WL contrast images shown in Fig.~\ref{fig:intro_figure}(j,k,l) to find the conductivity maps of the three graphene flakes. While the WL images only contain amplitude information~\cite{Jing2023}, we observe that the phase shift across the discrete measurement positions is in the range of 0.2-0.3~radians within the frequency range of the instrument, and we use the average value $\Delta\phi_\text{av}$ of this phase shift together with the amplitude ratio from the WL images to find the best estimate of the graphene conductivity based on numerical inversion of the FDM equations. The extracted conductivity is interpreted as the weighted conductivity across the 0.5--1.5~THz range, and the absolute values of the spatially resolved conductivity maps are shown in Fig.~\ref{fig:conductivity_maps}(a-c) for samples S1, S2, S3, respectively. As shown in the appendix (Fig.~\ref{am:lovat_complex_conductivity}), the real and imaginary parts of the conductivity maps show positive and negative values, consistent with the spectroscopy in Fig.~\ref{fig:conductivity_spectra}.  

\begin{figure}[h]
\centering
\includegraphics[width=0.9\textwidth]{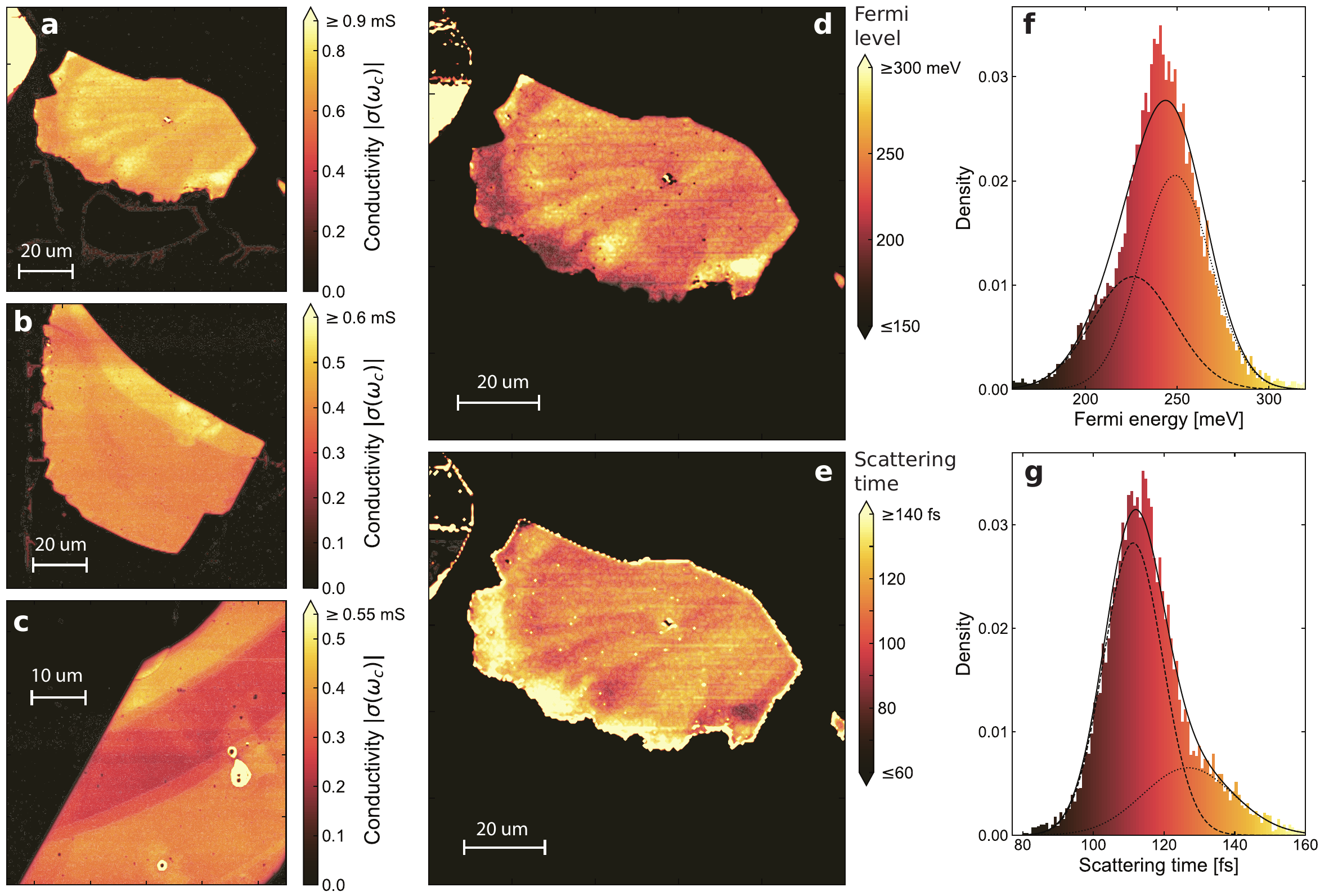}
\caption{$|$
\textbf{Conductivity and material parameter maps of three graphene samples.}
\textbf{a,b,c.} Absolute value of the complex conductivity extracted from white-light (WL) amplitude images combined with phase-resolved spectra at points P1–P4, averaged over 0.5–1.5~THz. Sample S1 (\textbf{a}) shows landscape-like variations with a pronounced low-conductivity region along the lower left edge. Sample S2 (\textbf{b}) exhibits generally lower, more uniform monolayer conductivity and enhanced bilayer conductivity. Sample S3 (\textbf{c}) clearly differentiates mono-, bi- and trilayer regions.  
\textbf{d,e.} Maps of Fermi energy $E_\mathrm{F}$ and scattering time $\tau$ for sample S1, obtained by inverting the Lovat-BGK analytical conductivity expression using both amplitude and average phase shift $\Delta\phi_\mathrm{av}$.   
\textbf{f,g.} Histograms of $E_\mathrm{F}$ and $\tau$ over the area in (\textbf{d,e}). Data are fitted with a double Gaussian distribution, where the main peak and shoulder correspond to the central flake region and the left-edge region, respectively.
}
\label{fig:conductivity_maps}
\end{figure}

The maps reveal a significant variation in the conductivity across the flakes. Figure~\ref{fig:conductivity_maps}(a) shows that the monolayer flake displays variations that visually resemble a height map of an eroded landscape, and we see a lower conductivity along the lower left edge of the sample. These variations will be discussed shortly. Sample S2 (Fig.~\ref{fig:conductivity_maps}(b)) shows an overall lower conductivity than sample S1, and a more uniform conductivity across the monolayer region. Furthermore, we observe a significantly higher conductivity of the bilayer region. Sample S3 (Fig.~\ref{fig:conductivity_maps}(c))) clearly shows the distinction between mono-, bi-, and trilayer graphene.  

For monolayer graphene (sample S1), we can use the conductivity maps to estimate the underlying physical parameters that determine the conductivity of the graphene layer. The Lovat formulation of the BGK conductivity of monolayer graphene depends on the Fermi energy, $E_\text{F}$, and the electron scattering time, $\tau$. Using the complex conductivity determined from the WL images and the spectral phase information, we can invert the Lovat equations to estimate the two material parameters (see Methods). Figure~\ref{fig:conductivity_maps}(d) shows the variations of the Fermi energy over the monolayer graphene flake, while Fig.~\ref{fig:conductivity_maps}(e) shows the variation of the scattering time. Based on these more fundamental maps, we can quantify the significant variation of the Fermi energy and scattering time. A pronounced anticorrelation between Fermi energy and scattering time is observed. While such behaviour is not expected for exfoliated graphene under ambient conditions, where long-range scattering dominates, it follows naturally from the structure of the conductivity model. Over the relevant regions of the flake, the conductivity exhibits only modest spatial variation, and since the intraband conductivity scales approximately with the product $E_\text{F}\tau$, increases in $E_\text{F}$ must be offset by decreases in $\tau$ in the inversion procedure. The significant variations in the material parameters across the flakes might be rooted in the transfer process, but this has not been explored in detail. However, inspection of several flakes (not shown here) indicates that the variations are not inherent properties of the mother graphite, as other flakes (including sample S2 and S3) show significantly smaller variations across monolayer regions. Independent Raman correlation analysis, shown in Appendix~\ref{amsec:raman}, yields Fermi energies consistent with the THz-SNOM-derived values.

The large number of data points provides statistical insight into variations in the electronic properties across the graphene flake. Figure~\ref{fig:conductivity_maps}(f,g) show histograms of the distribution of Fermi level (Fig.~\ref{fig:conductivity_maps}(f)) and scattering time (Fig.~\ref{fig:conductivity_maps}(g)). Both distributions are non-Gaussian. The Fermi level distribution shows a main peak at $\overline{E}_\text{F}=240\text{ meV}$ and a shoulder at lower energies (approx.~205~meV). The scattering time shows a peak at $\overline{\tau}=113\text{ fs}$, with a shoulder at higher scattering times (approx.~130-135~fs). The distributions can be fitted with a double-Gaussian distribution, as shown in Fig.~\ref{fig:conductivity_maps}(f,g) for the individual distributions and their sum. In spite of the significant overlap of the two distributions, they are indicative of the difference in electronic properties between the central part and the left edge of the flake. 

\section{Discussion}\label{sec:discussion}

The results presented here demonstrate direct experimental access to the momentum-dependent electrical conductivity $\sigma(\omega,q)$ of a two-dimensional conductor at nanometre length scales and terahertz frequencies. While nonlocal electronic response has long been predicted by microscopic transport theories and inferred indirectly in strongly confined plasmonic systems~\cite{Lindhard1954, Mermin1970, Kubo1957, Lundeberg2017, Aupiais2023}, it has remained largely inaccessible to direct measurement. By combining broadband terahertz near-field spectroscopy with quantitative electrodynamic modelling, we show that $\sigma(\omega,q)$ can be extracted experimentally rather than treated as a phenomenological correction to local response. This establishes nonlocal conductivity as a measurable material property in the terahertz regime.

A central implication of our findings is that commonly employed local-response models are insufficient for describing nanoscale electrodynamics at terahertz frequencies. Despite the use of AFM tips with radii of several hundred nanometres, superfocusing of the incident terahertz field generates in-plane momenta corresponding to length scales of order 50~nm~\cite{Cvitkovic2007, Fei2011}. At these scales, the conductivity of monolayer graphene is dominated by nonlocal response, leading to qualitative deviations from Drude-like behaviour, including a negative imaginary conductivity within the experimentally accessible frequency range~\cite{Lovat2013, Falkovsky2007}. This implies that modelling approaches based on spatially uniform conductivity or geometrically inferred characteristic momenta can yield qualitatively incorrect predictions of confinement, dispersion, and loss, even for device dimensions that are not conventionally considered at the extreme nanoscale.

In addition to quantitative near-field spectroscopy, our measurements reveal pronounced layer-dependent contrast in terahertz white-light near-field images. The clear contrast observed between mono-, bi- and trilayer graphene constitutes an important result in its own right. While near-field imaging of graphene at infrared and optical frequencies has previously revealed layer-dependent response primarily through interband transitions or plasmonic resonances~\cite{Lundeberg2017, Fei2011}. By contrast, earlier terahertz near-field imaging experiments found that graphene acts as an almost perfect reflector, yielding essentially no contrast between mono- and multilayer graphene\cite{Zhang2018}. The contrast observed here arises from a combination of improved measurement sensitivity and the use of dented probe tips, which increase the near-field signal and modify the characteristic in-plane momentum distribution sampled in the experiment.

In the terahertz regime, where photon energies are far below interband thresholds and conventional local models predict only weak layer dependence, the pronounced contrast observed here directly reflects differences in low-energy electronic response at the nanoscale. Our results, therefore, demonstrate that terahertz near-field microscopy is sensitive to subtle variations in electronic transport associated with layer number, even in the absence of spectrally resolved measurements for few-layer graphene.

We note that the terahertz near-field response observed here does not exhibit clear signatures of propagating surface plasmon–polariton modes. This is expected given the strong damping of graphene plasmons at terahertz frequencies ($\omega\tau\ll 1$) \cite{Anglhuber2025} and the irregular geometry of the exfoliated flakes, which precludes the formation of well-defined plasmonic cavities. Under these conditions, plasmon propagation lengths ($L/\lambda\approx\omega\tau/4\pi$) are comparable to or shorter than characteristic feature sizes, suppressing standing-wave interference and favouring a response predominantly determined by the graphene conductivity.

Beyond graphene, the near-field imaging and spectroscopic approaches demonstrated here are broadly applicable to low-dimensional conductors, where long carrier mean free paths coexist with deep-subwavelength field confinement. This includes a wide class of two-dimensional materials, oxide interfaces, and topological systems, where nonlocal transport is expected to influence plasmonic, electronic, and optoelectronic functionality. In such systems, terahertz near-field spectroscopy offers a unique route to experimentally benchmark microscopic transport models and to identify regimes where nonlocal response fundamentally constrains device performance. This is especially pertinent for topological materials, where conducting states may be confined to nanometre-scale regions and where direct access to local electrical response is otherwise largely limited to tunnelling probes of the local density of states or electrostatic surface-potential measurements.

In summary, this work shows that nonlocal electronic response is not a marginal correction but a dominant contribution to nanoscale conductivity in the terahertz regime. By directly measuring $\sigma(\omega,q)$ with nanometre spatial resolution, and by revealing pronounced layer-dependent contrast in terahertz near-field images, we establish terahertz near-field microscopy as a quantitative and versatile probe of nonlocal transport. These results provide a concrete experimental foundation for revisiting the definition, measurement, and modelling of conductivity at the interface between nanophotonics and high-speed electronics.

\section{Methods}\label{sec:methods}

\subsection*{Sample fabrication}
Graphene samples were prepared by mechanical exfoliation of graphite (NGS Naturgraphit GmbH) on silicon wafers with a 90~nm thick thermal SiO$_2$ layer, resulting in a wide range of graphene flakes with various thicknesses distributed on the same substrate. Optical microscopy was then used to identify suitable monolayer and few-layer flakes on the surface and to assess the number of graphene layers on a given sample based on the optical contrast~\cite{Casiraghi2007, Jessen2018}.

\subsection*{Setup for THz-SNOM measurements}
THz-SNOM is performed with a commercial s-SNOM system (Attocube THz-NeaSCOPE) equipped with an integrated THz time-domain spectroscopy module (Attocube/Menlo Systems TeraSmart). THz pulses are generated and detected by photoconductive antennas, and the integrated system has a useful signal-to-noise ratio in the frequency band 0.5--2~THz. The system operates as an atomic force microscope (AFM), with optical access to the AFM tip. The THz beam is guided and focused via reflective optics onto the AFM tip, and identical optics guide the scattered THz light from the AFM tip to the detector. We used solid PtIr AFM tips with a tip shank length of $80~\mu\text{m}$ (Rocky Mountain Nanotechnology, 25PtIr200B-H). We operated the THz-SNOM system in an atmosphere purged with nitrogen (N$_2$) gas to minimise water vapour absorption lines in the detected THz spectra. A schematic of the AFM tip on a sample with the incident and scattered THz radiation is shown in Fig.~\ref{fig:intro_figure}a.  

The THz-SNOM signal is recorded with the AFM operating in tapping mode, as is standard in most modern SNOM systems. The time-dependent scattered signal $S(t)$ is detected in real-time and demodulated at the overtones $S_m$ of the tapping frequency $\Omega$ to suppress the far-field background signal. Here $m$ is the demodulation order, typically $m=2-4$. Therefore, a time trace of a scattered THz signal is represented as $S_m(\tau)$, where $\tau$ is the time delay along the THz time axis, as shown in Fig.~\ref{fig:intro_figure}b. The frequency spectra of these two signals are shown in Fig.~\ref{fig:intro_figure}c, covering the 0.5--1.6~THz range.

\subsection{Analytical methods}

\subsubsection*{Finite Dipole Model (FDM) and multilayer model}

We use the Finite Dipole Model to describe the tip-sample interaction in our THz-SNOM analysis. The scattered electric field from the tip-sample system is modeled as $E_{\text{sca}} = (1+r_p)^2\alpha_{\text{eff}}E_{\text{inc}}$, where $E_\text{inc}$ is the incident field, $\alpha_\text{eff}$ is the effective polarizability of the tip-sample system, and $r_p$ is the far-field reflection coefficient of the incident $p$-polarized field. Due to the large dimensions of the tip and the focused THz spot, we ignore the spatial variations in $r_p$ in our analysis. The effective polarizability in the FDM is $\alpha_\text{eff}\propto1+\frac{1}{2}\beta f_0/(1-\beta f_i)$, with $\beta$ being the quasistatic near-field reflection coefficient and $f_0,i=\left(g-(R+2H+W_{0,i})/2L\right)\ln\left(4L/(R+4H+2W_{0,i})\right)/\ln(4L/R)$ are geometric factors determined by the tip-sample coupling $g\approx 0.7$, the tip major semiaxis length $L=40~\mu\text{m}$, $R$ is the tip radius (typically 200-400~nm), $W_0=1.31R, W_i=R/2$ are model locations of charges in the tip and substrate, respectively, and $H=H(t)=\frac{1}{2}A(1+\cos(\Omega t))$ being the distance from the tip apex to the sample surface. We calculate the demodulated signal at overtones of the tapping frequency as $E_\text{sca}(\Omega_m)=\int_0^TE_\text{sca}(t)e^{i\Omega_mt}dt$. 

In our experiment, the graphene samples are deposited on an SiO$_2$/Si substrate. Therefore, we need to incorporate the thin oxide layer in the modelling of the near-field reflection coefficient. We follow the matrix-based method formulated by Zhan \textit{et al.}~\cite{Zhan2013} and adopted for the description of the near-field reflection coefficient in the FDM by Wirth \textit{et al.}~\cite{Wirth2021}, where the reflection from a layered sample is modelled by $2\times 2$ matrices representing each interface and layer in the sample stack. Details are given in the Extended Data section. 

The in-plane momentum $q$ relevant for the evaluation of the near-field reflection coefficient in the FDM is calculated in the usual manner as described by Fei \textit{et al.}~\cite{Fei2011}. The weight function $q^2e^{-2qz_d}$ describes the weight of a given $q$ in the calculation of the tip-sample coupling. The time-averaged weight function is traditionally used as a reasonable approximation to the height-resolved tip-sample coupling, and with $z_d(t) = H(t)+R$, the time-average is $<q^2e^{-2qz_d}>=e^{-q(A+2R)}q^2I_0(Aq)$, where $I_0$ is the modified Bessel function of the first kind. This function has its maximum near $1/R$. 

\subsubsection*{White-light image processing}
White-light (WL) data contained only the scattering amplitude, with no phase information. Raw data are normalised to a 100-pixel area of the image recorded on the substrate, and the resulting scattering ratio amplitude $|\eta_m(x,y)|$ at demodulation orders $m=2,3$ are used for further analysis. 
WL data are recorded with the THz time delay parked at the peak of the THz transient, and are sensitive to any temporal shift of the THz signal during the scan~\cite{Jing2023}. We confirmed that the sample-induced shift is insignificant. The central frequency of the detected spectrum is approximately 1.2~THz, and the detected WL signal can be interpreted as a spectral average of the full waveform, as all frequencies in the bandwidth interfere constructively at the peak of the signal. We estimate an approximate phase of the WL signal by calculating the average phase from full-waveform measurements (see below). With the position-dependent amplitude and the constant phase estimate, we perform an inversion of the FDM model to retrieve a best estimate of the complex-valued conductivity map $\sigma_s(x,y)$. By inverting the nonlocal Lovat conductivity model at the weighted centre frequency of the THz spectrum, we then calculate the Fermi energy and scattering time maps $E_F(x,y)$ and $\tau(x,y)$. 

\subsubsection*{Spectroscopic analysis}
Full-waveform scattering signals $S_m(t)$ are Fourier transformed and normalised to substrate measurements to yield the frequency-dependent scattering ratio $\eta_m(\omega)$. With this input, we perform a numerical FDM inversion using a standard minimisation routine, minimising the distance estimate between experiment and model by variation of the complex-valued conductivity $\sigma_s(q,\omega)$ and the FDM model parameters tip radius $R$ and tapping amplitude $A$. For the extraction of Fermi energy $E_F$ and scattering time $\tau$ we then fit the Lovat model conductivity $\sigma_s(q,\omega)$ to the extracted conductivity, using the in-plane momentum in the Lovat model as a free parameter. The resulting best-fit value of $q$ is typically 5-10 times larger than the value inferred from the tip radius, consistent with the tight focusing of the field under the tip [Hillenbrand ref], see Additional Data section for further details. We use a scaling of the in-plane momentum between orders in the FDM model consistent with momentum-resolved simulations of the field distribution under the vibrating AFM tip [Huber ref].   

%%===========================================================================================%%
%% If you are submitting to one of the Nature Portfolio journals, using the eJP submission   %%
%% system, please include the references within the manuscript file itself. You may do this  %%
%% by copying the reference list from your .bbl file, paste it into the main manuscript .tex %%
%% file, and delete the associated \verb+\bibliography+ commands.                            %%
%%===========================================================================================%%

% TOGGLE-ON FOR EDITTING STATIC BIBLIOGRAPHY LIST
% \bibliography{THz-SNOM-bibliography}% common bib file

%%% STATIC BBL OUTPUT (TOGGLE FOR FINAL SUBMISSION - USE \bibliography to make edits)

%----------END OF BIBITEMS--------------

\begin{appendices}

\section{Extended data}\label{sec:extended}

\subsection{The Lovat model of nonlocal response in graphene}
\label{amsec:lovat_model}
%%An appendix contains supplementary information that is not an essential part of the text itself but which may be helpful in providing a more comprehensive understanding of the research problem or it is information that is too cumbersome to be included in the body of the paper.

Lovat et al.~published a model for the nonlocal response of graphene, valid for all in-plane momenta $q$~\cite{Lovat2013}. Their derivation is based on the Boltzmann transport equation. It enforces charge conservation by including charge diffusion as in the Bhatnagar-Gross-Krook (BGK) model in the same manner as the Mermin correction. Below is a reproduction of their analytical evaluation of the graphene conductivity tensor elements, with a relative error within 4\%~compared to their full expressions. 

\begin{eqnarray}
    \gamma &=& i\frac{e^2k_\text{B}T}{\pi^2\hbar^2}\ln\left\{2\left[1+\cosh\left(\frac{E_F}{k_\text{B}T}\right)\right] \right\}\ , \\
    \gamma_\text{D} &=& -i\frac{v_\text{F}}{2\pi\omega\tau}\ , \\
    \chi &=& \sqrt{1-\frac{v_\text{F}^2q^2}{\alpha^2}}\ , \\
    \sigma_\phi^\text{BGK} &=& \gamma\frac{2\pi\alpha}{v_\text{F}^2q^2}(1-\chi)\ , \\
    \sigma_\rho^\text{BGK} &=& \frac{v_\text{F}}{2\pi\gamma_\text{D}(1-\chi)-v_\text{F}\chi}\sigma_\phi^\text{BGK}\ ,
\end{eqnarray}
where $\alpha = \omega+i/\tau$ and $\sigma_\phi^\text{BGK}, \sigma_\rho^\text{BGK}$ are the transverse and longitudinal conductivities, respectively. 

Another widely used model for nonlocal conductivity is the hydrodynamic Drude model (HDM), in which the longitudinal conductivity can be written in the form
\begin{equation}
\sigma_\text{HDM} = \frac{\pi\gamma\omega}{\omega(\omega+i/\tau)-\beta_\text{eff}^2q^2}\ .   
\end{equation}
Here we have expressed the DC conductivity of graphene in terms of the Lovat notation ($\sigma_\text{DC}=-i\pi\gamma\tau$). In the simplest form of the HDM, $\beta_\text{eff} = \beta = v_F/\sqrt{2}$ for a 2D electron gas. This form of the HDM includes a force due to the electron pressure gradient. However, Landau damping is not included in the simplest form of the HDM. If a viscosity term $\eta\nabla^2\mathbf{v}$ is added to the HDM, $\eta$ being the dynamic shear viscosity, velocity gradients are damped and the nonlocal parameter becomes complex-valued with an imaginary part depending on frequency, $\beta_\text{eff} = \sqrt{v_F^2/2-i\omega\nu}$, where $\nu=\eta/(mn_0)$ is the kinematic viscosity estimated to be in the range $\nu=0.01-0.1\text{ m}^2/\text{s}$~\cite{Principi2016}. 

\begin{figure}[h]
\centering
\includegraphics[width=0.9\textwidth]{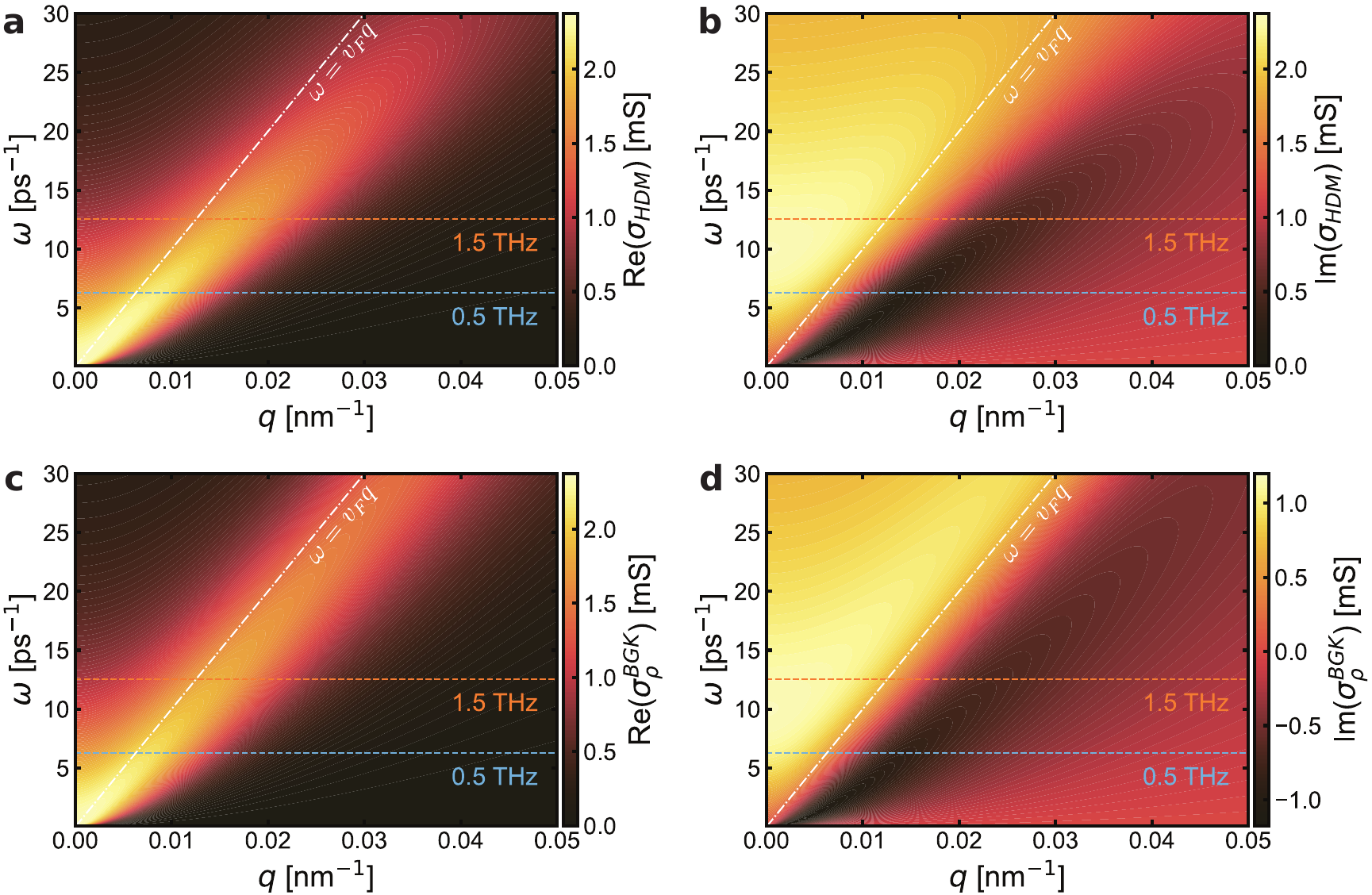}
\caption{Maps of nonlocal conductivity in the hydrodynamic Drude model (a,b: real and imaginary part, respectively) and in the Lovat formulation of the BGK model (c,d: real and imaginary part, respectively), plotted as a function of $q$ and $\omega$. Horizontal lines indicate frequencies 0.5 and 1.5~THz, see Fig.~\ref{am:lovat_vs_hdm_1d}. The diagonal dot-dashed line indicates $\omega=v_F q$ (see text).}\label{am:lovat_vs_hdm_2d}
\end{figure}

Figure~\ref{am:lovat_vs_hdm_2d} compares the longitudinal conductivity of the Lovat BGK model and the hydrodynamic Drude model, including the electronic viscosity term. Panels (a,b) in Fig.~\ref{am:lovat_vs_hdm_2d} display the real and imaginary parts of the conductivity in the hydrodynamic Drude model as a function of $q$ and $\omega$, using parameters $E_\text{F}=200\text{ meV}$, $\tau=100\text{ fs}$, $T=300\text{ K}$, $\beta=v_\text{F}/\sqrt{2}\approx 7.07\cdot 10^5\text{ m/s}$, and $\nu = 0.01\text{ m}^2\text{/s}$. For comparison, the Lovat BGK model conductivity is shown in Panels (c,d) in Fig.~\ref{am:lovat_vs_hdm_2d}, using the same parameters for graphene. Qualitatively, the two models display the same behaviour, with a resonant enhancement slightly below the graphene fermion dispersion line $\omega=v_\text{F}q$ (shown as a white dash-dotted line in each panel). The width of the resonant feature increases with frequency, indicative of Landau damping in the BGK model and electron viscosity in the HDM model.

\begin{figure}[h]
\centering
\includegraphics[width=0.9\textwidth]{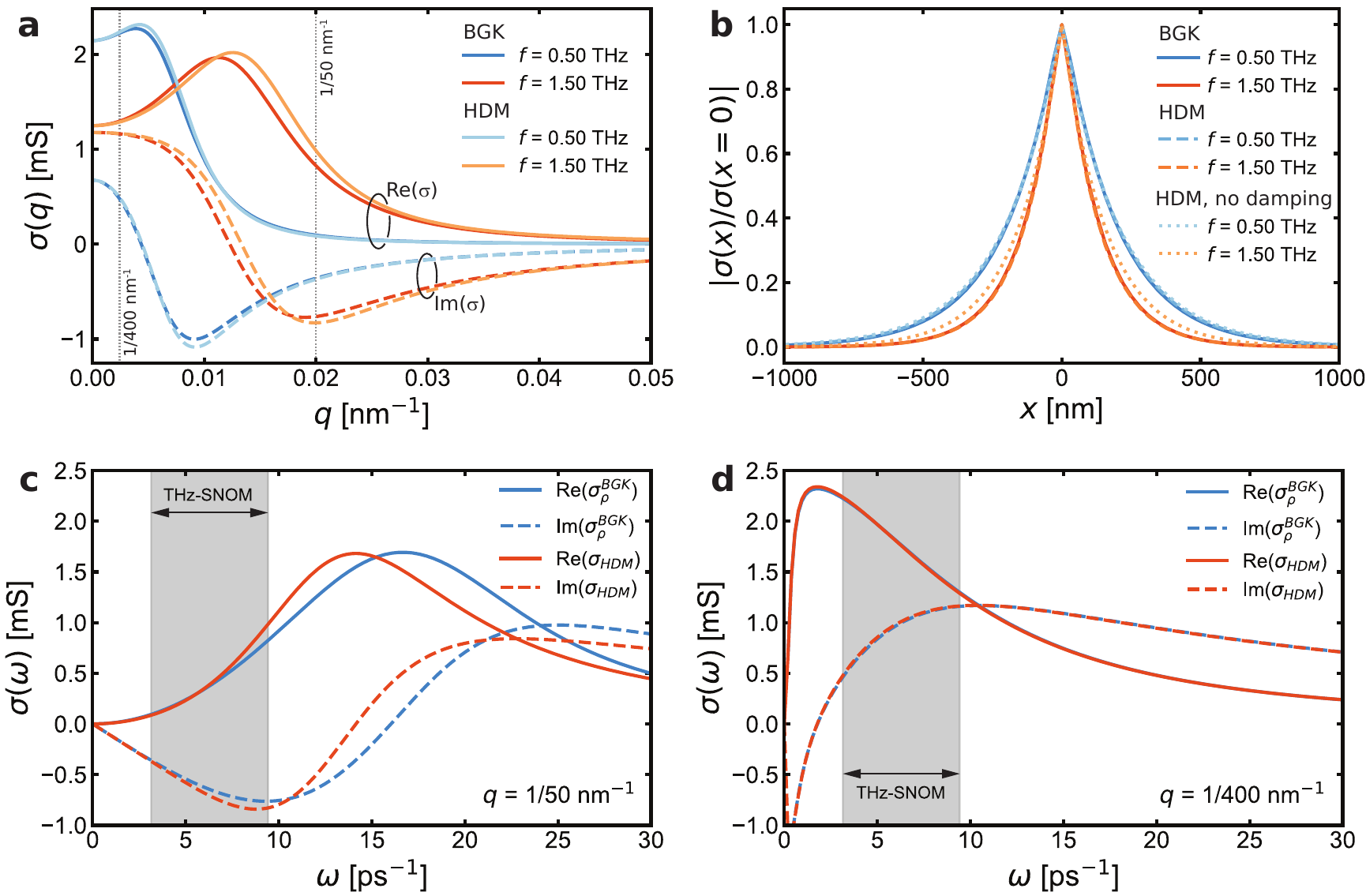}
\caption{(a) Momentum-dependent conductivity in the Lovat formulation of the BGK model and the HDM model at 0.5 and 1.5~THz (horizontal cross-sections indicated in Fig.~\ref{am:lovat_vs_hdm_2d}). (b) Real-space representation of the momentum-dependent BGK and HDM conductivity at 0.5 and 1.5~THz. The decay length indicates the nonlocal length scale. (c) Frequency-dependent conductivity at $q=1/50\text{~nm}^{-1}$, corresponding to the approximate localisation of the field under the AFM tip in THz-SNOM. (d) Same as (c), but at $q=1/400\text{~nm}^{-1}$, corresponding to the definition of the $R$ in the finite dipole model.}\label{am:lovat_vs_hdm_1d}
\end{figure}

Figure~\ref{am:lovat_vs_hdm_1d}(a) shows the momentum-resolved conductivity at $\omega/2\pi = 0.5$ and $1.5$~THz (the frequencies indicated by blue and orange dashed lines in Fig.~\ref{am:lovat_vs_hdm_2d}). The real and imaginary parts of the conductivities are shown as solid and dashed curves, respectively. Again, we see a reasonable agreement between the two models. Due to its simple mathematical form, the HDM can be used for a quantitative estimate of the spatial range of nonlocal effects. The inverse Fourier transform from momentum space to real space is
\begin{equation}
    \sigma_\text{HDM}(\omega,x) = -\frac{\gamma \pi^{3/2}\omega}{\sqrt{2}\beta_\text{eff}^2}\sqrt{-\frac{\beta_\text{eff}^2\tau}{\omega(i+\omega\tau)}}\exp\left(-|x|\sqrt{-\frac{\omega(i+\omega\tau)}{\beta_\text{eff}^2\tau}}\right)\ ,
\end{equation}
which, if we approximate $\beta_\text{eff}\approx\beta = v_F/\sqrt{2}$ (i.e. no viscous damping), leads to a $e^{-1}$ decay length of the nonlocality of
\begin{equation}
    L_\text{nl} = \frac{v_\text{F}}{\sqrt{\omega(\sqrt{\omega^2+1/\tau^2}-\omega)}}\ .
\end{equation}
Figure~\ref{am:lovat_vs_hdm_1d}(b) shows the spatial extent of the nonlocal character of the conductivity, as shown for the two models at the two selected frequencies 0.5 and 1.5~THz. The inverse Fourier transform of the Lovat BGK model is evaluated numerically, while the HDM curves represent the analytically evaluated transform. The dotted curves show the approximate analytical result $\sigma(x) \propto \exp(-|x|/L_\text{nl}$, which at the low frequency is very close to the full HDM expression, while a small deviation is seen at the higher frequency (1.5~THz), indicative of the frequency-dependent damping. The curves show that for practical purposes, the HDM and the more complete Lovat BGK models predict the same degree of nonlocality and momentum dispersion of the conductivity. 

Finally, Panels (c,d) of Fig.~\ref{am:lovat_vs_hdm_1d} display the frequency-dependent conductivity at two selected in-plane momenta ($q=1/50\text{~nm}^{-1}$ and $q=1/400\text{~nm}^{-1}$) of relevance to the current investigation. The greyed area indicates the spectral coverage of the THz-SNOM instrument. The conductivity evaluated at $q=1/50\text{~nm}^{-1}$ is relevant for the fitting of the extracted conductivity spectrum due to the tight confinement of the field under the tip to a radius significantly smaller than the geometrical dimensions of the tip (see Fig.~\ref{fig:nonlocality} in the main text), and we see a close resemblance between the modeled and the measured conductivity spectra (compare with Fig.~\ref{fig:conductivity_spectra}). On the other hand, for $q=1/400\text{~nm}^{-1}$, which is close to the weighted in-plane momentum based on the geometrical tip radius and the tapping cycle relevant for the inversion of the quasi-electrostatic Finite Dipole Model, the conductivity spectrum is relatively close to that of the local Drude model except at the lowest frequencies below the THz-SNOM coverage. This conductivity spectrum is not observed in our measurements, again indicating that the field is tightly confined relative to the physical tip radius.  

\subsection{Multilayer model for the near-field reflection coefficient}
\label{amsec:multilayer_model}

The matrices $\tensor{D}_{12}$ and $\tensor{D}_{23}$ are the transfer matrices for $p$-polarized light across the interface between air and SiO$_2$ and the SiO$_2$/Si interface, respectively, 
\begin{eqnarray}
    \tensor{D}_{12} = \frac{1}{2}\begin{bmatrix} 1+\eta_{12}+\xi_{12} & 1-\eta_{12}-\xi_{12} \\ 1-\eta_{12}+\xi_{12} & 1+\eta_{12}-\xi_{12} \end{bmatrix} \ , \ \ 
    & & \tensor{D}_{23} = \frac{1}{2}\begin{bmatrix} 1+\eta_{23} & 1-\eta_{23} \\ 1-\eta_{23} & 1+\eta_{23} \end{bmatrix} \ , \nonumber \\
    \eta_{12} = \frac{\epsilon_1k_{2z}}{\epsilon_2k_{1z}} \ , \ \ \ \ \xi_{12} = \frac{\sigma_s k_{2z}}{\epsilon_0\epsilon_2\omega}\ , & & \ \ \eta_{23} = \frac{\epsilon_2k_{3z}}{\epsilon_3k_{2z}} \ .
\end{eqnarray}
Here $k_{iz} = \sqrt{\epsilon_ik_0^2-q^2}$ is the out-of-plane wave number in medium $i$ and $\epsilon_i$ is the permittivity of medium $i$ ($i=1$: air, $i=2$: SiO$_2$, $i=3$: Si). 

The propagation matrix through the SiO$_2$ spacer layer of thickness $d_{\text{SiO}_2}$ is
\begin{equation}
    \tensor{P}(\Delta z) = \begin{bmatrix} e^{-ik_z\Delta z} & 0 \\ 0 & e^{ik_z \Delta z}\end{bmatrix}\ .
\end{equation}
The full transfer matrix of the layered structure is then $\tensor{M} = \tensor{D}_{12}\tensor{P}\tensor{D}_{23}$. For a conductive film directly on a semi-infinite substrate, $\tensor{M} = \tensor{D}_{12}$. In both cases, the near-field reflection coefficient $\beta(\omega,q)$ for the Finite Dipole Model is then extracted from the elements of $\tensor{M}$,
\begin{equation}
    \beta(\omega,q) = \frac{M_{21}}{M_{11}}\ .
\end{equation}

\subsection{Fitting parameter overview}
\label{amsec:fitting_parameters}

Here we reproduce the fitting parameters for the FDM model and the conductivity models used to fit the experimental data in Fig~\ref{fig:conductivity_spectra} of the main article. The tip radius $R$ in the FDM is used as a fitting parameter, and the conductivity model has parameters $E_\text{F}$, $\tau$, and $q$ as fit parameters. We note that the in-plane momentum used in the fits to the Lovat conductivity model differs from the dominant in-plane momentum used in the FDM, as discussed in the main article. We used a tapping amplitude of $A = 100\text{~nm}$ in all fitting procedures. We observed that the final fitted parameters were insensitive to the precise value of $A$, so we fixed it to improve the stability of the fitting routines. 

\subsubsection{Time-domain fitting procedures}

\begin{table}[h]
    \centering
    \begin{tabular}{|c|ccccc|}
    \hline
    \cellcolor[HTML]{c7dbee}Position & \cellcolor[HTML]{c7dbee}$E_\text{F}$ [meV] & \cellcolor[HTML]{c7dbee}$\tau$ [fs] & \cellcolor[HTML]{c7dbee}$q$ [nm$^{-1}$] & \cellcolor[HTML]{c7dbee}$R$ [nm] & \cellcolor[HTML]{c7dbee}$\Delta$ \\ \hline    
    P1 & 241.87 & 75.36 & 0.0212 & 370.93 & 0.0508 \\ 
    P2 & 232.62 & 56.77 & 0.0198 & 349.43 & 0.0487 \\
    P3 & 293.45 & 60.74 & 0.0224 & 273.80 & 0.0615 \\
    P4 & 203.55 & 66.85 & 0.0194 & 257.43 & 0.0592 \\ \hline
    \end{tabular}
    \caption{Best-fit parameters, sample S1, time-domain minimisation, Lovat conductivity model.}
    \label{tab:S1_TD}
\end{table}

\begin{table}[h]
    \centering
    \begin{tabular}{|c|ccccc|}
    \hline
    \cellcolor[HTML]{c7dbee}Position & \cellcolor[HTML]{c7dbee}$E_\text{F}$ [meV] & \cellcolor[HTML]{c7dbee}$\tau$ [fs] & \cellcolor[HTML]{c7dbee}$q$ [nm$^{-1}$] & \cellcolor[HTML]{c7dbee}$R$ [nm] & \cellcolor[HTML]{c7dbee}$\Delta$ \\ \hline    
    P1 & 500$^\dagger$ & 8.19 & 0.0017$^{\dagger\dagger}$ & 400$^\star$ & 0.2623 \\ 
    P2 & 500$^\dagger$ & 10.52 & 0.0017$^{\dagger\dagger}$ & 400$^\star$ & 0.2506 \\
    P3 & 500$^\dagger$ & 8.99 & 0.0017$^{\dagger\dagger}$ & 400$^\star$ & 0.2650 \\
    P4 & 500$^\dagger$ & 8.78 & 0.0017$^{\dagger\dagger}$ & 400$^\star$ & 0.2502 \\ \hline
    \end{tabular}
    \caption{Best-fit parameters, sample S1, time-domain minimisation, Drude conductivity model. $\dagger$ - upper limit of fit range; $\dagger\dagger$ - lower limit of fit range; $\star$ - fixed parameter.}
    \label{tab:S1_TD}
\end{table}

\begin{table}[h]
    \centering
    \begin{tabular}{|c|ccccc|}
        \hline
        \cellcolor[HTML]{c7dbee}Position &\cellcolor[HTML]{c7dbee}$E_\text{F}$ [meV] &\cellcolor[HTML]{c7dbee} $\tau$ [fs] &\cellcolor[HTML]{c7dbee} $q$ [nm$^{-1}$] &\cellcolor[HTML]{c7dbee} $R$ [nm] &\cellcolor[HTML]{c7dbee} $\Delta$ \\ \hline
        P1 & 347.30 & 50.00$^\dagger$ & 0.0258 & 312.90$^\star$ & 0.00148 \\
        P2 & 271.92 & 50.00$^\dagger$ & 0.0216 & 312.90$^\star$ & 0.00152 \\
        P3 & 379.52 & 50.00$^\dagger$ & 0.0269 & 312.90$^\star$ & 0.00176 \\
        P4 & 294.45 & 50.00$^\dagger$ & 0.0238 & 312.90$^\star$ & 0.00157 \\ \hline
    \end{tabular}
    \caption{Best-fit parameters, sample S1, frequency-domain minimisation, Lovat conductivity model. $\dagger$ - lower limit of fix range; $\star$ - fixed parameter, average of time-domain fit results (see Table~\ref{tab:S1_TD}).}
    \label{tab:S1_FD}
\end{table}

\subsubsection{Frequency-domain fitting procedures}

\begin{table}[h]
    \centering
    \begin{tabular}{|c|ccccc|}
    \hline
    \cellcolor[HTML]{fed1a3}Position & \cellcolor[HTML]{fed1a3}$E_\text{F}$ [meV] & \cellcolor[HTML]{fed1a3}$\tau$ [fs] & \cellcolor[HTML]{fed1a3}$q$ [nm$^{-1}$] & \cellcolor[HTML]{fed1a3}$R$ [nm] & \cellcolor[HTML]{fed1a3}$\Delta$ \\ \hline
    P1 & 151.23 & 51.05 & 0.0179 & 402.90 & 0.0433 \\
    P2 & 151.53 & 50.00$^\dagger$ & 0.0192 & 368.09 & 0.0396 \\
    P3 & 150.00 & 50.00$^\dagger$ & 0.0187 & 361.53 & 0.0370 \\
    P4 & 171.87 & 50.00$^\dagger$ & 0.0195 & 372.22 & 0.0404 \\ \hline
    \end{tabular}
    \caption{Best-fit parameters, sample S2, time-domain minimisation, Lovat conductivity model. $\dagger$ - lower limit of fit range.}
    \label{tab:S2_TD}
\end{table}

\begin{table}[h]
    \centering
    \begin{tabular}{|c|ccccc|}
        \hline
        \cellcolor[HTML]{fed1a3}Position &\cellcolor[HTML]{fed1a3}$E_\text{F}$ [meV] &\cellcolor[HTML]{fed1a3} $\tau$ [fs] &\cellcolor[HTML]{fed1a3} $q$ [nm$^{-1}$] &\cellcolor[HTML]{fed1a3} $R$ [nm] &\cellcolor[HTML]{fed1a3} $\Delta$ \\ \hline
        P1 & 152.81 & 50.00$^\dagger$ & 0.0181 & 376.19$^\star$ & 0.00101 \\
        P2 & 157.79 & 50.00$^\dagger$ & 0.0194 & 376.19$^\star$ & 0.00094 \\
        P3 & 162.30 & 50.00$^\dagger$ & 0.0188 & 376.19$^\star$ & 0.00090 \\
        P4 & 88.39 & 138.97 & 0.0163 & 376.19$^\star$ & 0.00126 \\ \hline
    \end{tabular}
    \caption{Best-fit parameters, sample S1, frequency-domain minimisation, Lovat conductivity model. $\dagger$ - lower limit of fit range; $\star$ - fixed parameter, average of time-domain fit results (see Table~\ref{tab:S2_TD}).}
    \label{tab:S3_FD}
\end{table}

\subsubsection{Gaussian fits to Fermi level and scattering time distributions}

The distributions of estimated Fermi levels and scattering times in Fig.~\ref{fig:conductivity_maps}(f,g) in the main paper are fitted with a Gaussian Mixture Model (GMM), giving maximum‐likelihood estimates for weights, means, and variances via the Expectation-minimisation algorithm. Here, we summarise the fit results used to produce the fitting curves shown in the figure of the main article. The parameters are
\begin{equation}
    f(x) = W_1 \exp\left(-\frac{(x-\mu_1)^2}{\sigma_1^2}\right)+W_2 \exp\left(-\frac{(x-\mu_2)^2}{\sigma_2^2}\right)\ ,
\end{equation}
with weight factors $W_1,W_2$, means $\mu_1,\mu_2$, and standard deviations $\sigma_1,\sigma_2$ for the two peaks. 
\begin{table}[h]
    \centering
    \begin{tabular}{|c|cc|}
        \hline
        \cellcolor[HTML]{fed1a3}Parameter & \cellcolor[HTML]{fed1a3}$E_\text{F}$ distribution &\cellcolor[HTML]{fed1a3}$\tau$ distribution \\ \hline
        $W_1$ & 0.38 & 0.74 \\
        $W_2$ & 0.62 & 0.26 \\ 
        $\mu_1$ & 226~meV & 111~fs \\
        $\mu_2$ & 249~meV & 127~fs \\ 
        $\sigma_1$ & 22.4~meV & 8.3~fs \\ 
        $\sigma_2$ & 19.3~meV & 12.8~fs \\ \hline
    \end{tabular}
    \caption{Best-fit parameters, GMM fitting to Fermi level and scattering time distributions.}
    \label{tab:S4_gaussians}
\end{table}

\subsection{Sensitivity analysis}
\label{amsec:sensitivity_analysis}

\begin{figure}[ht]
  \centering
  \includegraphics[width=\linewidth]{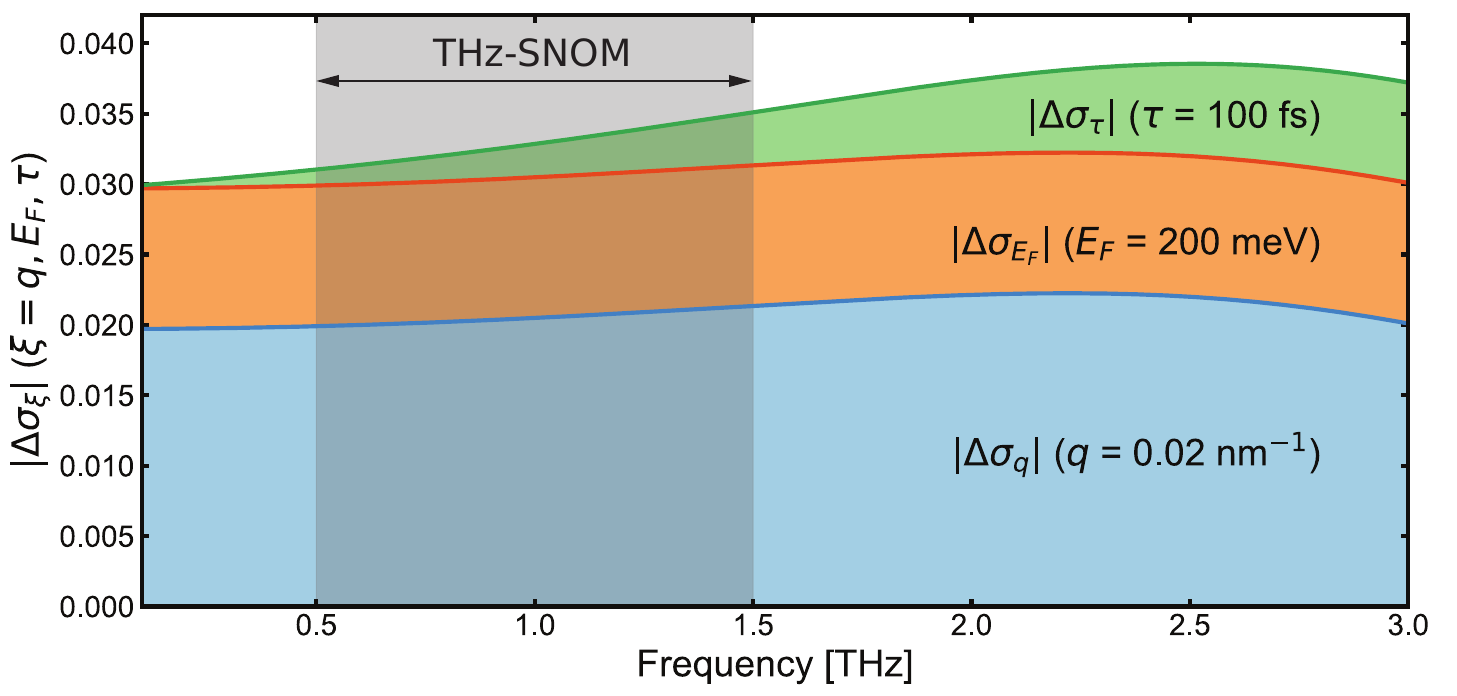}
  \caption{Sensitivity of the Lovat-model graphene conductivity to 1\% variations in in-plane momentum $q$, Fermi energy $E_\text{F}$, and scattering time $\tau$. The plotted quantity $\lvert\Delta\sigma_\xi\rvert = \lvert\sigma^{-1}\partial\sigma/\partial\xi\rvert\,\Delta\xi$ is evaluated at $q=0.02$~nm$^{-1}$, $E_\text{F}=200$~meV, and $\tau=100$~fs. The grey band indicates the frequency range of our THz-SNOM system (0.5--1.5~THz), within which variations in $q$ dominate the conductivity response.}
  \label{am:lovat_sensitivity}
\end{figure}

The BGK conductivity of graphene in the Lovat model is governed by three parameters: the Fermi energy $E_\text{F}$, the carrier scattering time $\tau$, and the in-plane momentum $q$. When fitting this model to our THz-SNOM data, we found that variations in $\tau$ have minimal impact on the fit quality. To quantify this, Fig.~\ref{am:lovat_sensitivity} shows the relative sensitivity
\begin{equation}
    |\Delta\sigma_\xi|=\left|\frac{1}{\sigma}\frac{\partial\sigma}{\partial\xi}\right|\Delta\xi\ ,
\end{equation}
for a 1\%~perturbation in each parameter ($\xi=q,E_\text{F},\tau$) evaluated at $q=0.02$~nm$^{-1}$, $E_\text{F}=200$~meV, and $\tau=100$~fs, across the 0.5--1.5~THz range (shaded). Within this band, a 1\%~change in $q$ produces roughly a 2\%~change in $\sigma$, whereas a 1\%~change in $E_\text{F}$ yields only about a 1\%~change, and $\tau$ contributes negligibly. Hence, the measured conductivity spectrum is predominantly set by the in-plane momentum.

\subsection{Complex-valued conductivity maps}
\label{amsec:complex_conductivity_maps}

In the main article, we show the maps of the absolute value of the conductivity across the three samples S1, S2, S3 (Fig.~\ref{fig:conductivity_maps}). For completeness, we here show the real and imaginary components of the conductivity maps. Similar to the spectrally resolved point measurements (Fig.~\ref{fig:conductivity_spectra} in the main article), we find a positive real part and a negative imaginary part in the maps, consistent with a nonlocal response.
\begin{figure}[ht]
  \centering
  \includegraphics[width=\linewidth]{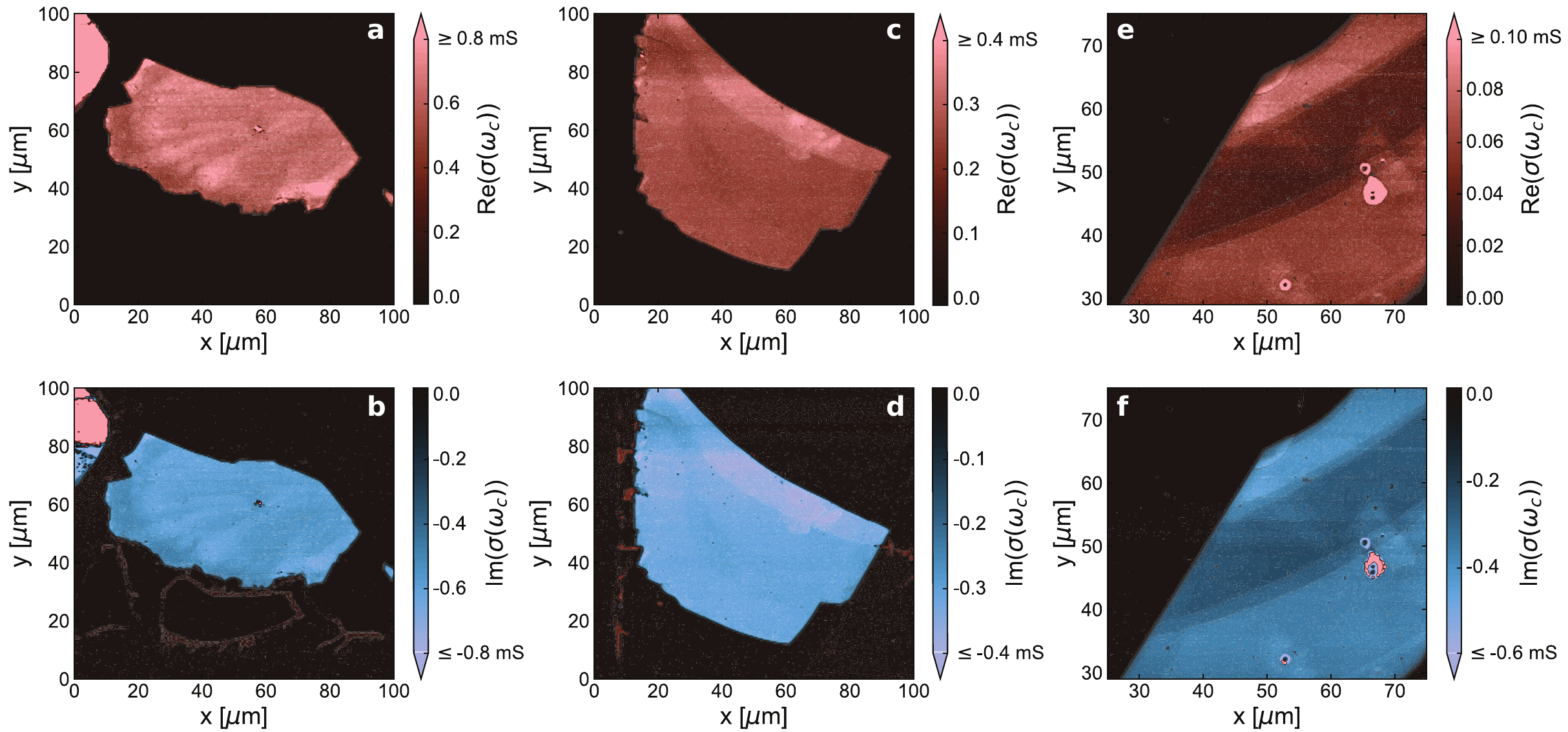}
  \caption{Real and imaginary parts of the conductivity maps of the graphene flakes. (a,b) Sample S1, (c,d) sample S2, (e,f) sample S3.}
  \label{am:lovat_complex_conductivity}
\end{figure}

\subsection{Raman correlation analysis}
\label{amsec:raman}
\begin{figure}[ht!]
  \centering
  \includegraphics[width=\textwidth]{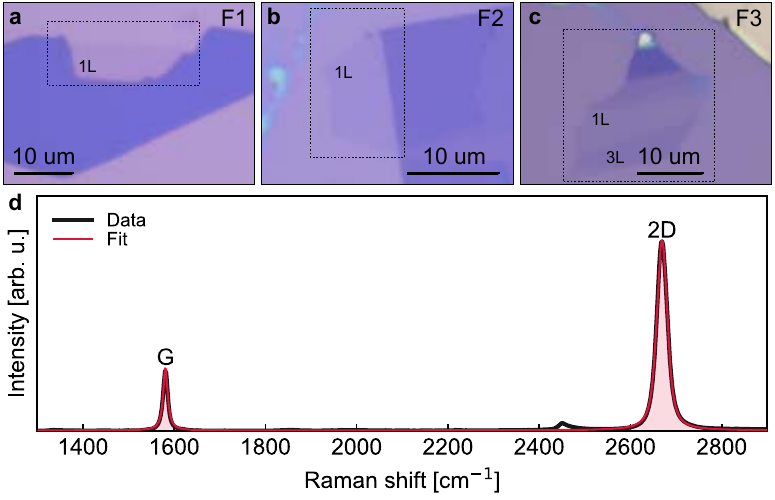}
  \caption{Optical microscopy and representative Raman spectroscopy. (a-c), Optical micrographs of three distinct monolayer graphene flakes, denoted F1, F2 and F3, prepared for spectroscopic characterisation.
  Dashed boxes identify the regions selected for high-resolution two-dimensional Raman mapping, which are predominantly monolayer (1L).
  Scale bars, $10~\mu$m.
  (d) Spatially-averaged Raman spectrum (black) corresponding to the monolayer region of the flake F1 (highlighted in (a)).
  The experimental data is overlaid with a cumulative numerical fit (red), utilising a Lorentzian profile for the G-peak and a Voigt profile for the 2D-peak.
  The negligible intensity of the defect-activated D-band ($\sim$1350 cm$^{-1}$) confirms the high structural integrity of the exfoliated flake.}
  \label{am:composite_raman}
\end{figure}
\begin{figure}[ht!]
  \centering
  \includegraphics[width=\textwidth]{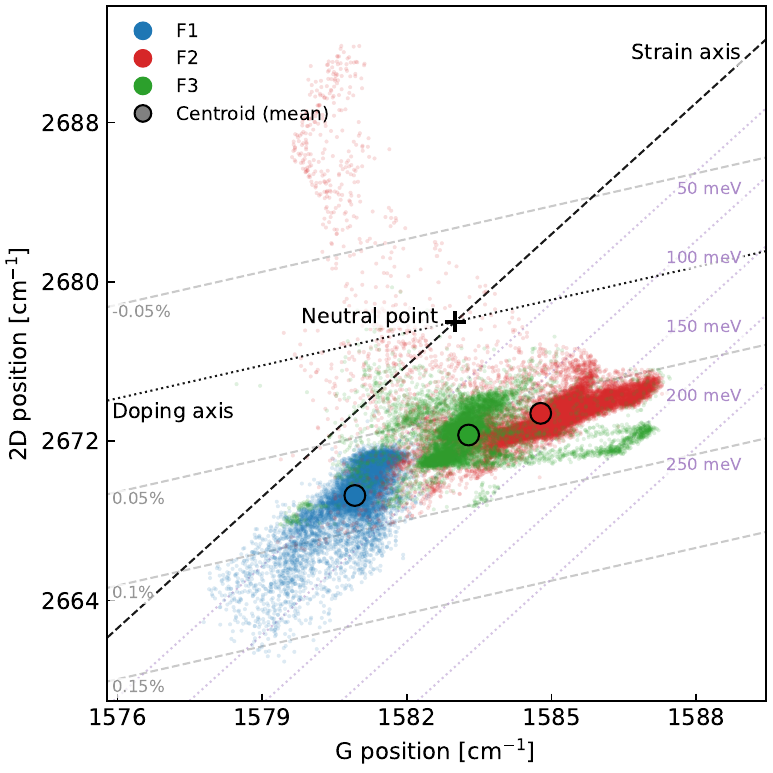}
  \caption{Raman correlation map of $\omega_{2D}$ versus $\omega_G$ peak positions for isolated monolayer regions on three distinct graphene flakes (F1, blue; F2, red; F3, green). 
  The black cross marks the theoretical charge neutrality point for 532 nm excitation.
  The black dashed and dotted lines define the non-orthogonal basis vectors for pure strain (slope $\sim$2.2) and pure doping (slope $\sim$0.55), respectively, while the lighter grid lines indicate isolines for constant strain (grey, dashed) and Fermi energy (purple, dotted). 
  Large circles denote the centroids of each distribution.
  The elongation of the F1 cluster along the strain axis reveals moderate substrate-induced tensile variation while maintaining a low intrinsic doping level ($E_\mathrm{F} < 100 $ meV).
  The clusters for flakes F2 and F3 align more closely with the doping trajectory, exhibiting variation up to moderate doping levels ($E_\mathrm{F} < 200 $ meV). 
  These data corroborate the THz-SNOM findings by attributing the lower Fermi energies observed here to the pristine nature of freshly exfoliated flakes, in contrast to the moderately aged samples used in the SNOM study.}
  \label{am:raman_correlation}
\end{figure}

Raman characterisation was performed on graphene samples fabricated using the identical protocol to the SNOM study, ensuring comparable material quality and substrate conditions.
Optical micrographs of the three mapped flakes (F1-F3) are presented in Fig.~\ref{am:composite_raman} (a-c).
Fig.~\ref{am:composite_raman} (d) shows a representative spectrum averaged over the monolayer region of F1, overlaid with a cumulative numerical fit (red), employing a Lorentzian profile for the G-peak and a Voigt profile for the 2D-peak. 

To quantify the electronic and mechanical properties, we employed the vector decomposition method initially formulated by Lee~\textit{et al.}~\cite{Lee2012} and adapted for substrate-supported graphene by Mueller~\textit{et al.}~\cite{Mueller2018} and Vincent~\textit{et al.}~\cite{Vincent2019}.
This technique separates strain and doping by analysing the correlation between the G and 2D mode frequencies ($\omega_\mathrm{G}$ and $\omega_\mathrm{2D}$).
The resulting correlation map is shown in Fig.~\ref{am:raman_correlation} for the three mapped flakes: F1, F2 and F3.
In the correlation space, frequency shifts due to strain and doping evolve along distinct, linearly independent trajectories. 
The experimental peak positions can thus be expressed as a vector sum relative to a charge-neutral, strain-free reference point ($\omega^0_\mathrm{G}$, $\omega^0_\mathrm{2D}$):
\begin{equation}
\begin{pmatrix} 
\omega_\mathrm{G} - \omega_\mathrm{G}^0 \\ 
\omega_\mathrm{2D} - \omega_\mathrm{2D}^0 
\end{pmatrix} 
= \Delta\omega_{\varepsilon_\mathrm{s}} \mathbf{v}_{\varepsilon_\mathrm{s}} + \Delta\omega_{n} \mathbf{v}_{n},
\label{eq:raman_vector}
\end{equation}
where $\Delta\omega_{\varepsilon_\mathrm{s}}$ and $\Delta\omega_{n}$ represent the frequency shifts projected along the strain and doping axes, respectively, and $\hat{\mathbf{v}}_{\varepsilon_\mathrm{s},n}$ represents the corresponding non-orthogonal basis vectors. 
For graphene supported on SiO$_2$, we adopt the reference point $\omega^0_\mathrm{G} = 1583$~cm$^{-1}$ and $\omega^0_\mathrm{2D} = 2678$~cm$^{-1}$, which are the established charge-neutrality values for 532~nm excitation of unencapsulated graphene~\cite{Mueller2018, Das2008}. 
The slope of the strain axis was set to $\Delta\omega_\mathrm{2D}/\Delta\omega_\mathrm{G} = 2.2$~\cite{Lee2012}. 
For the doping axis, we utilized a slope of $\Delta\omega_\mathrm{2D}/\Delta\omega_\mathrm{G} = 0.55$, reflecting the suppression of the 2D mode stiffening due to substrate dielectric screening~\cite{Vincent2019, Mueller2018, Froehlicher2015}, distinct from the value of 0.7 observed in suspended graphene~\cite{Lee2012}.

Solving the linear system in Eq.~\ref{eq:raman_vector} isolates the doping-induced component of the G-peak shift.
The Fermi energy was calculated using the relation $|E_\mathrm{F}| = |\Delta\omega_\mathrm{G}^\mathrm{dop}|/S_\mathrm{G}$, where $S_\mathrm{G} \approx 42$~cm$^{-1}$/eV. 
This sensitivity factor is adopted from detailed Raman studies of high-quality graphene~\cite{Froehlicher2015}, which refined earlier experimental estimates~\cite{Das2008} to match theoretical predictions for p-type doping~\cite{Lazzeri2006}.

The Raman analysis confirms the robustness of the fabrication protocol, placing the graphene consistently in a low-to-moderate doping regime ($E_\mathrm{F}<200$ meV).
The F1 cluster elongates primarily along the strain axis, indicative of residual substrate-induced tensile strain and low doping, whereas the F2 and F3 clusters show variation along the doping trajectory up to moderate levels ($E_\mathrm{F} <200$ meV). 
These results establish the intrinsic baseline of the fresh monolayers and corroborate our THz-SNOM measurements, attributing the slightly higher Fermi energies observed in that study to moderate sample ageing.

\end{appendices}

\end{document}